\documentclass[reprint,amsmath,amssymb,aps]{revtex4-2}

\usepackage{graphicx}
\usepackage{dcolumn}
\usepackage{bm}
\usepackage{hyperref}
\usepackage{graphicx,amssymb,amsmath,amsfonts}
\usepackage{verbatim}

\def\be{\begin{equation}}
\def\ee{\end{equation}}

\newcommand{\Rea}{\mathrm{Re}}
\newcommand{\Ima}{\mathrm{Im}}
\newcommand{\avg}[1]{\langle #1 \rangle}

\begin{document}

\title{A measure for chaotic scattering amplitudes}

\author{Massimo Bianchi}
 \email{massimo.bianchi@roma2.infn.it}

\author{Maurizio Firrotta}
 \email{maurizio.firrotta@gmail.com} 
\affiliation{Dipartimento di Fisica, Università di Roma Tor Vergata,\\
Via della Ricerca Scientifica 1, 00133, Roma, Italy
}
\affiliation{INFN sezione di Roma Tor Vergata \\
Via della Ricerca Scientifica 1, 00133 Roma, Italy
}

\author{Jacob Sonnenschein}
\email{cobi@tauex.tau.ac.il}
\affiliation{The Raymond and Beverly Sackler School of Physics and Astronomy, \\
Tel Aviv University, Ramat Aviv 69978, Tel Aviv, Israel}

\affiliation{Simons Center for Geometry and Physics, SUNY, Stony Brook, NY 11794, USA}

\author{Dorin Weissman}
\email{dorin.weissman@oist.jp}
\affiliation{Okinawa Institute of Science and Technology \\1919-1 Tancha, Onna, Okinawa 904-0495, Japan}

\date{\today}
\begin{abstract}
We propose a novel measure of chaotic scattering amplitudes. It takes the form of a log-normal distribution function for the ratios $r_n={\delta_n}/{\delta_{n+1}}$ of (consecutive) spacings $\delta_n$ between two (consecutive) peaks of the scattering amplitude.
We show that the same measure applies to the quantum mechanical scattering on a leaky torus as well as to the decay of highly excited string  states into two tachyons. Quite remarkably the $r_n$ obey the same distribution that governs the non-trivial zeros of Riemann zeta function.
\end{abstract}

\maketitle



\section{Introduction}
Chaotic behavior in classical and quantum mechanical systems has been intensively researched. The study of ergodicity and chaos in continuous quantum field theories and string theories is less mature.

Recently it was proposed in \cite{Rosenhaus:2020tmv,Gross:2021gsj,Rosenhaus:2021xhm} to investigate chaotic behavior in string theory by analyzing amplitudes of highly excited string states. The amplitude that was computed is a three-point function: the decay of an excited string into two tachyons. The calculation is done in open bosonic string theory in flat spacetime in the critical dimension \(D=26\).

The erratic behavior in the string amplitude of \cite{Gross:2021gsj,Rosenhaus:2021xhm} was demonstrated by the apparent large differences between plots of the decay amplitudes of two states with excitation number $N$ that differ only slightly in the partitioning of $N$. Two natural questions that arise are: what is the origin of this type of chaotic behaviour and how to quantify it in this system -- or in any other scattering processes. We will show that the chaotic behavior of string amplitudes can be associated to the large fluctuations on the spin of the highly excited state, even with fixed mass, and propose a new measure of chaos associated with these processes. This measure should be applicable to a wide range of scattering problems.

In order to motivate our proposal, let us recall a common method used to diagnose chaos in Hamiltonian systems, whose random observables are the spacings between energy levels \cite{Berry:1977wk},
\begin{equation} \delta_n = E_{n+1} - E_{n} \label{eq:defdelta} \end{equation}
or the \emph{ratios of successive spacings}
\begin{equation} \label{eq:rn} r_n \equiv \frac{E_{n+1}-E_n}{E_n-E_{n-1}} = \frac{\delta_{n+1}}{\delta_n} \end{equation}
In chaotic systems the level spacings are distributed as the spacings of eigenvalues of random matrices. The related distribution function for $r_n$ has also been successfully used as a measure of chaos \cite{Huse,Srdinsek:2020bpq} \footnote{In some cases the normalized ratio \(\tilde r_n \equiv min\{r_n,\frac{1}{r_n}\}\), defined to be between 0 and 1, is used.}. 

In analogy to the energy level spacings and the ratio $r_n$, we propose to analyze the scattering amplitude ${\cal A(\alpha)}$ of \cite{Gross:2021gsj,Rosenhaus:2021xhm} using the spacings between successive peaks of ${\cal A}$ as a function of $\alpha$, where $\alpha$ is a continuous kinematical variable of the scattering problem.

For our present purposes we define the spacings \(\delta_n\) between peaks and the ratios $r_n$ as in equations \eqref{eq:defdelta} and \eqref{eq:rn}, but replacing the energy levels $E_n$ with $\alpha_n$, the positions of peaks of ${\cal A}(\alpha)$. Our main result is the analysis of the decay amplitude of a highly excited open bosonic string state into two tachyons. In this case the continuous variable $\alpha$ is the difference between the angle of the outgoing tachyons and the photons used to create the initial state in the approach of Del Giudice, Di Vecchia and Fubini (DDF), first introduced in \cite{DelGiudice:1971yjh} (see also \cite{Hindmarsh:2010if, Skliros:2011si, Bianchi:2019ywd, Addazi:2020obs, Aldi:2019osr} for reviews and modern applications). The spacings are erratic in the sense that they follow the same patterns as the spectra of chaotic systems.

A prototype system displaying chaotic behavior is given by the positions of the non-trivial zeros of the Riemann zeta function \cite{Berry:1986,Odlyzko:1987,Wolf:2014ulr}. The normalized spacings between successive zeros were long observed to follow a distribution very close to the random matrix theory result for the Gaussian unitary ensemble (GUE).

The chaotic properties of the zeta function are directly relevant to the physical problem of scattering on the ``leaky torus'', which is a torus with an infinite cusp \cite{Gutzwiller:1983}. The analytic result for the phase shift \(\Phi\) of an incoming wave as a function of its momentum \(k\), which is the continuous variable relevant for this case, has an erratic part that is essentially the phase of \(\zeta(s)\) on the line \({\rm Re}[s]=1\). As such, when we look at the spacings for successive maxima of the function \(\Phi(k)\), the phase displays similar patterns as the non-trivial zeros on the critical line \({\rm Re}[s] = \frac12\).\footnote{A relation to string amplitudes is also found in the representation of the Veneziano amplitude in terms of zeta functions \cite{He:2015jla,CastroPerelman:2022ypp}. However, the Veneziano amplitude is certainly not chaotic, so further steps are necessary to link chaos in the zeta function to that of string amplitudes.}

While the distributions of the spacings differ for the three problems we examined, we find that they all exhibit a simple, smooth distribution for the spacing ratios \(r_n\) that can be fitted well by a log-normal distribution. That is, \(\log r\) is normally distributed, and the probability distribution function (PDF) for \(r\) is
\begin{equation} f_{\text{LN}}(r) = \frac{1}{\sqrt{2\pi \sigma^2} r} \exp\left(-\frac{(\log r-\mu)^2}{2\sigma^2}\right) \label{eq:pdfLN} \end{equation}
The random matrix theory distribution for the GUE is very close to a log-normal distribution, with an appropriate choice of \(\sigma\). The distributions we find are all symmetric in \(r\to1/r\), setting \(\mu=0\), and all have a value of \(\sigma\) within 10\% or less of the value closest to the GUE. 

The definition of \(r_n\) depends on the ordering of the spacings. If the spacings \(\delta_n\) are all independent random variables, then their order should not matter. Given the set of spacings $\{\delta_n\} = \{\delta_1,\delta_2,\delta_3,\ldots\}$ one can perform the ``shuffling test'': changing the order of the spacings by considering instead the set $\{\delta_{\sigma(n)}\}$ where $\sigma(n)$ is some permutation of the indices (which can be either random or predetermined), and then checking if the distribution of \(r_n\) remains invariant. The caveat is that one needs to properly normalize the spacings to remove any overall, average \(n\)-dependence from \(\delta_n\).

\section{Chaos in the Riemann zeta function}
We begin with the analysis of the zeros of the zeta function, since in this case we can benefit from a wealth of available data. We use it as an archetypal example to compare with the following analysis of the leaky torus phase shift and the bosonic string amplitudes, and show that the log-normal distribution for the spacing ratios is an effective model.

Let us recall some well known results. The Riemann hypothesis states that all non-trivial zeros of the zeta function are located on the ``critical line'' \(\Rea[s] = \frac12\) \footnote{As opposed to the trivial zeros located at \(s = -2n\) for integer \(n\).}, and are therefore of the form
\begin{equation} s_n = \frac12 + i z_n \end{equation}
with real \(z_n\) \footnote{Since \(\zeta(s^*) = \zeta^*(s)\) we can discuss zeros with positive imaginary part only.}.

Further analysis shows that the normalized spacings \footnote{The logarithmic normalization is not essential in this case. One can normalize also by dividing by the mean, \(\bar \delta_n \equiv \delta_n/\avg{\delta}\). In both cases \(\bar \delta=1\).},
\begin{equation} \bar \delta_n \equiv {z_n-z_{n-1}\over 2\pi}\log{z_n\over 2\pi} \label{eq:deltaNorm} \end{equation}
follow the GUE distribution whose PDF is \footnote{More precisely, we write the PDF for $2\times2$ random matrices, which is generally considered an excellent approximation of the PDF of spacings for large matrices as well.}
\begin{equation} p_{\text{GUE}}(\bar \delta) = \frac{32}{\pi^2} \bar\delta^2 \exp\left(-\frac{4\bar \delta^2}{\pi}\right) \label{eq:pdfGUE} \end{equation}
With the assumption that the spacings \(\bar \delta_n\) are independent random variables drawn from \eqref{eq:pdfGUE}, then the PDF of their ratios \(r_n\) is predicted to be
\begin{equation} f_{\text{GUE}}(r) = \frac{16}{\pi} \frac{r^2}{(1+r^2)^3} \label{eq:pdfGUE_r}\end{equation}
This function is peaked at \(r = 1/\sqrt2\), with the expectation value of \(r\) being \(\avg{r} = 4/\pi \approx 1.273\). This distribution turns out to be very similar to the log-normal distribution with \(\mu = 0\), and \(\sigma = \sqrt{(\pi/2)^2-2)}\) \footnote{This specific value of \(\sigma\) is the one for which the relative entropy, \(\int f_{\text{GUE}}\log(f_{\text{GUE}}/f_{\text{LN}})\), is minimized, with the minimum being \(\approx 0.008\).}.



\begin{figure}[ht!]
    \centering
    \includegraphics[width=0.30\textwidth]{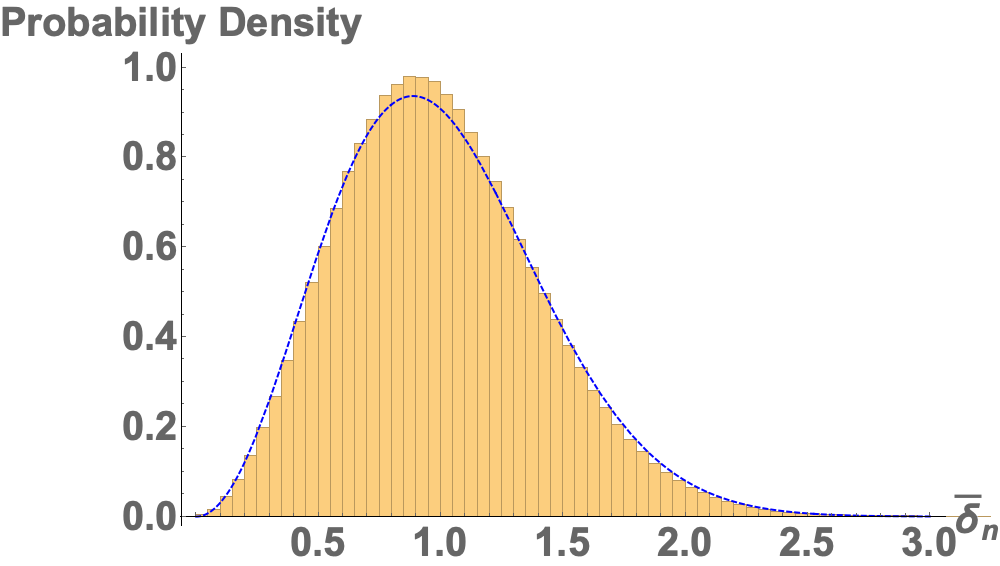}
    \includegraphics[width=0.30\textwidth]{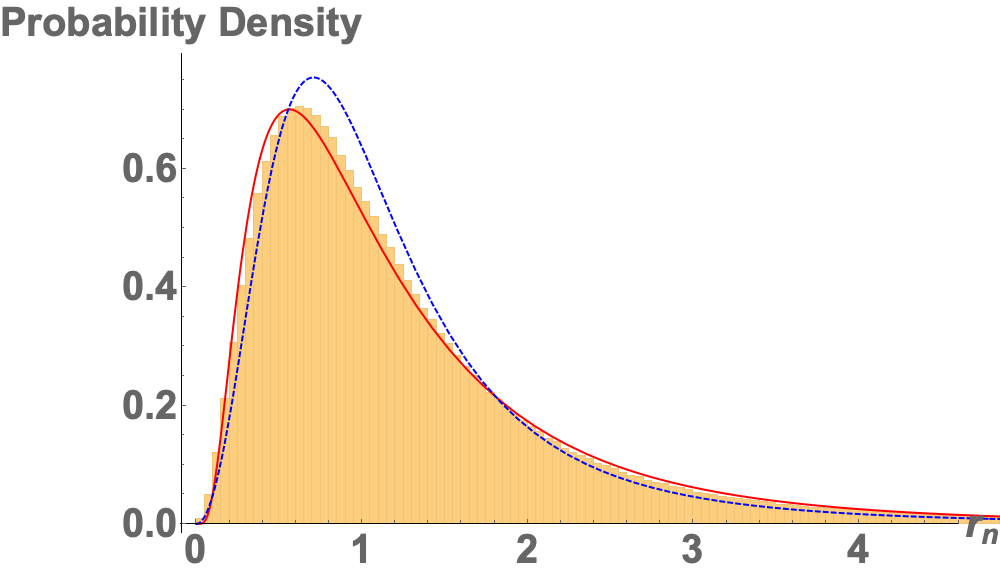}
    \caption{Distributions of \(\bar\delta_n\) (top) and \(r_n\) (bottom) for zeros of the zeta function. The dashed blue lines are the GUE prediction, and, in the bottom plot, we add the log-normal fit in red.}
    \label{fig:zeta_zero}
\end{figure}

We use the table of zeros of the zeta function available online at \cite{Odlyzko:Zeta} for the first \(N = \) 2,001,052 zeros. The distribution of the normalized spacings and their ratios are drawn in figure \ref{fig:zeta_zero}. The distributions are very well fitted by the GUE predictions $p_{GUE}(\bar\delta)$ and $f_{GUE}(r)$. When we fit the distribution of the ratios to a log-normal distribution $f_{LN}(r)$ we find better agreement, though this might be expected with an additional free parameter. There is a noticeable difference between the average measured value \(\avg{r_n}_N = 1.347\) and the \(4/\pi\approx1.273\) predicted by eq. \eqref{eq:pdfGUE_r}. Interestingly, a random shuffling of the order of spacings results in a distribution of \(r\) much closer to the GUE one, and brings the average value down to \(\avg{r}_{\text{shuf.}} \approx1.25\), implying that there are correlations between neighboring spacings, which the shuffling eliminates.

The average value of \(r_n\) increases slowly when the range \(z_N\) is increased. It appears to converge to some value around \(1.35\) but a persistent yet mild logarithmic growth cannot be ruled out from the data.

\section{Chaos of the zeta function in scattering: The leaky torus}
The geometry of the leaky torus \cite{Gutzwiller:1983} is constructed as follows. In the two dimensional hyperbolic space with the metric \footnote{We implicitly set the intrinsic length scale in the problem to one when writing the metric.}
\begin{equation} ds^2 = \frac{dx^2+dy^2}{y^2} \label{eq:leaky_metric} \end{equation}
one looks at the region, in the upper half plane \(y>0\), between the geodesics (i) \(x=-1\), (ii) \(x = 1\), (iii) \((x-\frac12)^2+y^2 = (\frac12)^2\), and (iv) \((x+\frac12)^2+y^2 = (\frac12)^2\).
Then, identifying boundary (i) with (iii) and (ii) with (iv), the result is a torus with a cusp point at infinity.

The ``scattering experiment'' one can perform in this geometry is to send a free wave from \(y=\infty\) and measure the phase shift between incoming and outgoing wave at some finite \(y = y_0 > 0\). There is a remarkable analytic result for this phase shift $\Phi(k)$ as a function of the momentum of the incoming wave \(k\) \cite{Gutzwiller:1983}. The chaotic behavior comes from a term which is simply the phase of the Riemann zeta function \(\zeta(s)\) along the line \(\Rea[s]=1\). Namely, one finds
\begin{equation} \Phi(k) = \frac{\Ima[\zeta(1+2 i k)]}{\Rea[\zeta(1+2 i k)]} \label{eq:leaky_function} \end{equation}
The erratic behavior is already apparent when plotting \(\Phi(k)\) as a function of \(k\). To analyze it further we study the distribution of local extrema. We look at extrema instead of zeros to make the analysis more like the one for the string amplitude in the following section.

In the present analysis \(\{z_n\}\) are the zeros of the derivative, i.e. points satisfying $\Phi^\prime(z_n) = 0$. These can be either local minima or maxima. We collected the first \(N =\) 22,618 zeros of \(\Phi^\prime(k)\). The first is at \(k_1\approx 3.19\), and the last in our set is \(k_N \approx \) 12,927. Half of the points are local maxima and half are minima. We have three options: looking at all extrema, only maxima, or only minima. The distributions have the same or similar shape, but the one for all extrema has a larger variance of both spacings and spacing ratios, because of occasional points where a minimum and a maximum are very close to each other, which does not occur for minima or maxima separately.

\begin{figure}[t!]
    \centering
        \includegraphics[width=0.30\textwidth]{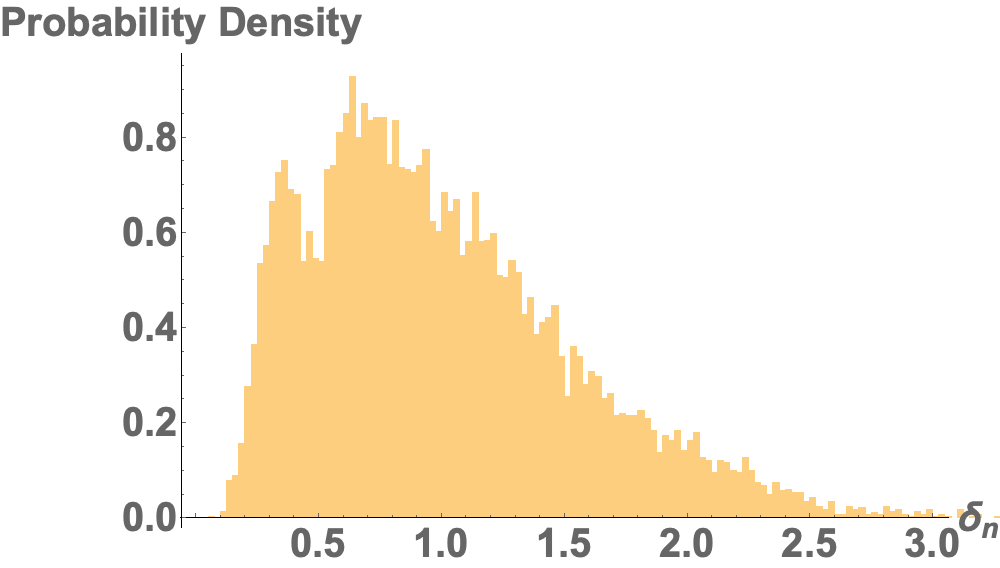} \\
    \includegraphics[width=0.30\textwidth]{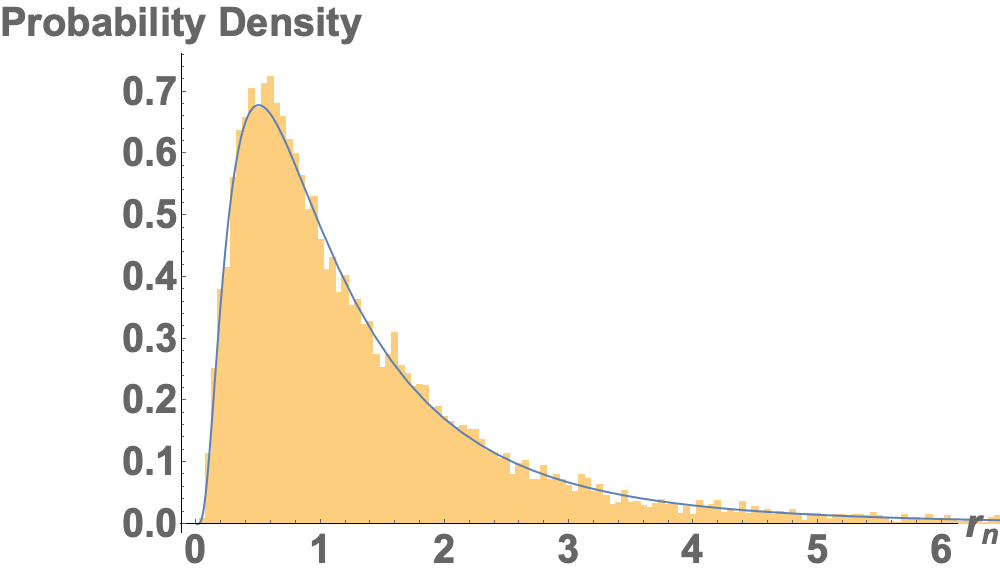}
    \caption{Distribution of normalized spacings \(\delta_n/\avg{\delta_n}\) (top) and their ratios (bottom) for the location of the first 11,309 maxima of \eqref{eq:leaky_function}.}
    \label{fig:leaky_r}
\end{figure}

We normalize the spacings by dividing by the mean value, \(\bar \delta_n \equiv \delta_n/\avg{\delta}\). There is a difference between the distributions of spacings compared with the case of zeros of the zeta function. Here the underlying distribution of \(\delta_n\) is more complex and appears to have more than one peak. However, the distribution of spacing ratios is again well modeled as a log-normal distribution. We plot the distribution of spacings of maximum points in figure \ref{fig:leaky_r}. The average values we find are \(\avg{r}_{\text{min}} = 1.394\), \(\avg{r}_{\text{max}} = 1.418\), \(\avg{r}_{\text{all}} = 1.944\). The distributions for minima and maxima turn out to be similar to that of the zeta function zeros, being log-normal with \(\mu\approx0\) and a very similar average value.

\section{Chaos in amplitudes of an excited string}
\subsection{Amplitude of a highly excited state and two tachyons}
Let us now consider the three point function of an excited string state \(H_N\) and two tachyons in open bosonic string theory. The study of chaotic behavior in this amplitude was initiated in \cite{Gross:2021gsj,Rosenhaus:2021xhm}, with details of the derivation to be found mainly in \cite{Gross:2021gsj}. The only reliable way of constructing BRST invariant vertex operators for highly excited states and computing their amplitudes relies on the DDF approach \cite{DelGiudice:1971yjh, Hindmarsh:2010if, Skliros:2011si, Bianchi:2019ywd}. In the standard covariant approach, identifying BRST invariant vertex operators becomes extremely challenging already at relatively low levels \cite{Bianchi:2010dy, Bianchi:2010es}, except for the first Regge trajectory $J=N$ \cite{Bianchi:2011se,Black:2011ep,Schlotterer:2010kk}.  The latter do not produce any chaotic behavior as shown in \cite{Aldi:2020qfu, Aldi:2021zhh, Firrotta:2022cku}.
 
The definition of the DDF operators \(A_n\) involves a reference null momentum $q$, and a set of circular polarizations $\lambda_n$ transverse to $q$. Taking all $\lambda_n$ to be equal simplifies the analysis while still exposing the chaotic behavior. For generic random choices of  $\lambda^i_n$ the chaos would be even more evident, at the cost of significantly more involved calculations.

Thus, the state \(H_N\) to be considered is
\begin{equation} |H_N\rangle = \prod_{m=1}^\infty \left(\lambda \cdot A_m\right)^{n_m} |0\rangle \label{eq:HNi}\end{equation}
and is defined by an integer partition \(\{n_m\}\) for which
\begin{equation} N = \sum_{m=1}^\infty m\, n_m\,, \qquad J = \sum_{m=1}^\infty n_m \end{equation}
It is important to note that despite the notation, which we retain from \cite{Rosenhaus:2021xhm}, \(J\) is not the total spin of the state, but rather its helicity. Had \(H_N\) been a state of definite spin, the angular distribution would have been completely determined by the spin and displayed no erratic behavior. The angular distribution becomes unpredictable and erratic because the states \(H_N(\{n_m\})\) are complex superpositions of many states of different spin \(s\geq J\).

The erratic function which we will analyze is the angle dependence of the amplitude for \(H_N\) to decay to two tachyons. At lowest (disk) order one finds
\begin{equation} {\cal A}_{H_N\to TT} \sim (\sin \alpha)^J  \prod_{m=1}^\infty \left[\sin(\pi m \cos^2\frac\alpha2)\right]^{n_m} \label{eq:A_Rosenhaus} \end{equation}
The angle \(\alpha\) is the angle between the outgoing tachyons and the photons used to create the DDF state.
For computation and visualization it is preferable to use the logarithmic derivative,
\begin{align*} F(\alpha) \equiv& \frac{d}{d\alpha}\log {\cal A} =\\ \nonumber =& J \cot \alpha - \frac\pi2\sin\alpha \sum_{m=1}^\infty m n_m \left[\cot(\pi m \cos^2\frac\alpha2)\right]^{n_m} \label{eq:dlogA}\end{align*}
This is an erratic function that can have many zeros, which are the peaks of the amplitude \eqref{eq:A_Rosenhaus}.\footnote{More precisely, in terms of the $|{\cal A}|$ there are only maxima. In this sense, unlike the case of the leaky torus, all the zeros of the derivative represent peaks.} As we show momentarily these peaks are randomly spaced.

\subsection{Statistics}
For a given state, we define \(\{z_n\}\) as the set of zeros of \(F(\alpha)\) in the range from 0 to \(\pi\). For these zeros, as before, we define the spacings and their ratios \(r_n\).

The set of \(\{r_n\}\) is defined for a particular state \(H_N\). We will study the distribution of values in the union of many such sets for many different states, keeping only \(N\) fixed.

The number of peaks of the amplitude, or zeros of \(F(\alpha)\), depends on the state considered. This number scales linearly in \(N\), which is also the maximal possible spin of a state at that level. For very large \(N\) we can gather enough data points for a statistical analysis from a single amplitude. For very small \(N\), there are not enough points for a meaningful analysis. For intermediate \(N\), a smooth distribution of spacings and their ratios emerges when we accumulate data points from many different states at the same level \(N\).

The total number of states of the form \eqref{eq:HNi} at level \(N\) with fixed \(\lambda\) is equal to the number of integer partitions of \(N\), which for large \(N\) grows as
\begin{equation} p_N \approx \frac{1}{4\sqrt{3} N} e^{C \sqrt{N}}\,, \qquad C \equiv \pi \sqrt{\frac23}\label{eq:partitions} \end{equation}
The fraction of partitions of \(N\) which have fixed length \(J\) is approximately given by a Gumbel distribution \cite{Erdos:1941}, i.e. according to the PDF
\begin{equation} d_N(J) = \frac{1}{\beta} \exp\left(-\frac{J-\mu}{\beta} - e^{-\frac{J-\mu}{\beta}}\right) \label{eq:Gumbel} \end{equation}
with the parameters scaling as \(\mu \sim \sqrt{N} \log N\) and \(\beta \sim \sqrt{N}\) \footnote{More precisely the distribution is peaked around \( J \approx \frac{1}{C} \sqrt{N}\log N\) with the same constant \(C\) as in \eqref{eq:partitions}.}.

Amplitudes with \(J \approx N\) are not chaotic. The case \(J=N\), the leading Regge trajectory, is not chaotic at all as it is a single state with definite spin at each level. The function \({\cal A}(\alpha)\) has a single peak in this case. However such states become an exponentially small fraction of the states, and based on \eqref{eq:Gumbel} we can impose the condition \(N > \frac{1}C\sqrt{N}\log N\) that assures that most states are far from the end-point \(J \approx N\) that represents the leading Regge trajectory.

For intermediate \(N\), such as \(N = 100\), it is not practical to analyze sets of many millions or more of states, and we can proceed by taking a small sample of states as follows. For a given \(N\), we draw a sample of 5000 values of \(J\) from the distribution of eq. \eqref{eq:Gumbel}. Then, given the list of \(J\), for each value in the list we draw a random partition of \(N\) into \(J\) integers. In this way we end up with 5000 randomly selected states. Then, calculating the spacings for each state in the sample, and taking the union of all such sets in our sample should result in a subset of the spacings that is distributed in the same way as the set of spacings for all amplitudes at the level \(N\), for the reason that the helicities of the states we used follow the same distribution as all the states at that level.

\begin{figure}[pt!]
    \centering
    \includegraphics[width=0.32\textwidth]{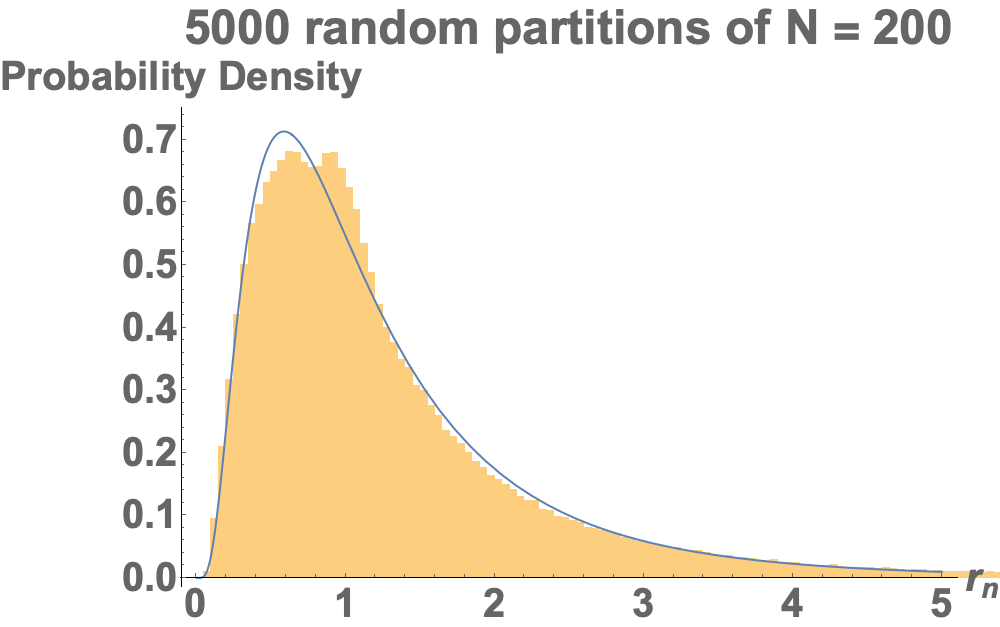}
    \includegraphics[width=0.32\textwidth]{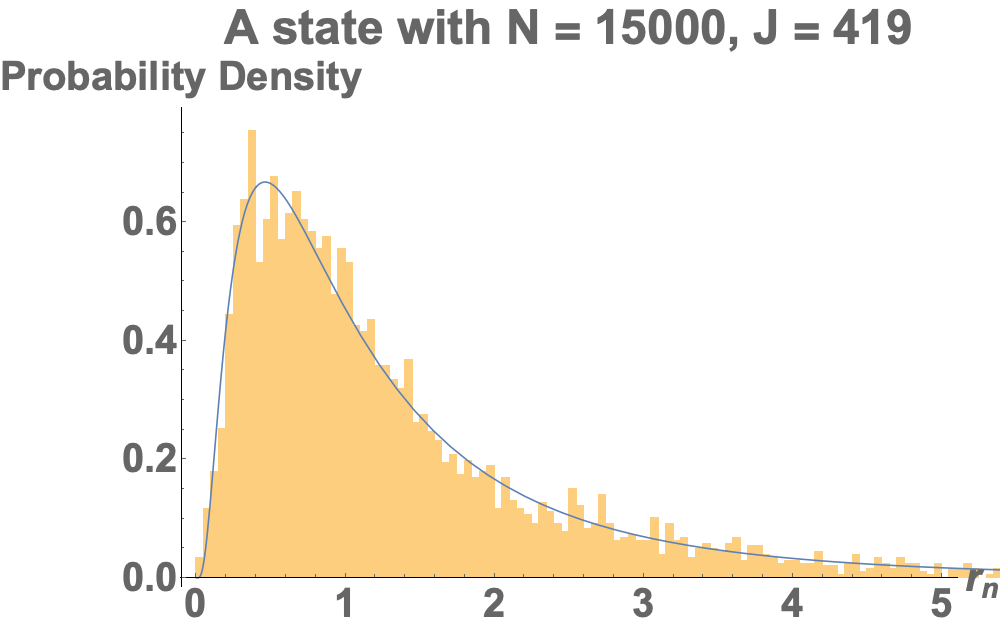}
    \caption{Combined (aggregate) distributions of \(\{r_n\}\) for many states with \(N =  200\) (top) and distribution of \(\{r_n\}\) for a single state of \(N = \)15,000 (bottom).}
    \label{fig:rString}
\end{figure}

The result is always that the distribution of many values of \(r_n\) is well approximated by a log-normal distribution. We draw a representative example in figure \ref{fig:rString}. Table \ref{tab:results} summarizes our results, including the dependence on \(N\) and \(J\) across the different samples taken.

\begin{table}[h!]
    \centering
    \begin{tabular}{|l|l|l|l|c|c|} \hline
         Level  &   No. of states  &  Sampled & Total & Per & Average \\
          &                         &  states   & points    & state     & \(\langle r_n \rangle\)\\
         \hline\hline
         \(N = 15\)  & 176 &   176  & 982 & 6 & 1.059 \\ \hline
         
         \(N = 20\)  & 627 &   627  & 4,923 & 8 & 1.079 \\ \hline
         
         \(N = 25\) &   1958 &   1958  & 19,947 & 10 & 1.097 \\ \hline
         
         \(N = 30\) &  5604  &  5604  & 69,791 & 12 & 1.114 \\ \hline
         
         \(N = 50\) & \(\approx\)204,226  & 5000 & 103,535 & 20 &  1.173 \\ \hline
         
          \(N = 100\) & \(\approx1.9\times 10^8\)  & 5000 & 201,470 & 40 & 1.247 \\ \hline
         
         \(N = 100\)  &\(\approx1.1\times 10^7\)  & 2000 & 88,360 & 44 & 1.236 \\ 
         \(J = 18\)  &  & &  &  &  \\ \hline
         
         \(N = 100\)  & 1958   & 1958 & 25,420 & 14 & 1.313 \\ 
         \(J = 75\)  &  & &  &  &  \\ \hline
         
         \(N = 200\) & \(\approx4\times 10^{12}\)  & 5000 & 383,764 & 76 &  1.307 \\ \hline
         
          \(N = 200\)  &\(\approx1.6\times10^8\)  & 2000 &  169,040 & 84 & 1.294 \\ 
         \(J = 28\)  &  & &  &  &  \\ \hline
         
          \(N = 200\)  & 1958  & 1958 &  25,420 & 14 & 1.348\\ 
         \(J = 175\)  &  & &  &  &  \\ \hline

         \(N = \)10,000  & \(\sim 10^{106}\)  & 20 & 52,669 & 2,578 & 1.522 \\ 
         
         \hline
         
    \end{tabular}
    \caption{States used in the analysis and results. Total data points is the total number of values of \(r_n\) in a given sample, and the number of data points per state is a median value.}
    \label{tab:results}
\end{table}

One observation from the table is that the average \(\avg{r_n}\) increases slowly with \(N\). Whether it converges or continues growing as \(N\) is taken to infinity cannot be determined from the data. This is similar to the dependence of the average in the two previous examples on increasing the range of zeros.

The log-normal distribution is a good approximation overall, but one can see from the plots that the data has an excess of points around \(r = 1\) compared to the fitted distribution. It is not clear what is the source. The excess goes away if one shuffles the spacings, suggesting that there is some correlation in the data between neighboring spacings, \(\delta_n\) and \(\delta_{n+1}\). These could also be artifacts of the particular way in which we chose our states, or of an approximation taken en route to the simple form of the amplitude in eq. \eqref{eq:A_Rosenhaus} \footnote{The form of each term \(\sin(\pi m \cos^2\frac\alpha 2)\) is obtained from a ratio of \(\Gamma\) functions, after a Stirling approximation assuming \(m\) is large, which is not always the case.}.

\section{Summary}
We proposed a new measure of chaotic behavior of scattering processes. We used it to analyze three particular systems, but we expect that it could be used for a wide range of scattering problems in QFT and string theory. 

We have shown that there is a simple, smooth distribution of the ratios of spacings in the string amplitudes for highly excited open bosonic strings, which we model as a log-normal distribution. If one would carry out the experiment of taking a generic highly excited string and measuring the angular distribution in its decay to two tachyons, the result will be essentially unpredictable, despite the simple form of the string amplitude \eqref{eq:A_Rosenhaus}.

A skeptic could claim that this is not a result of string dynamics, but an artifact of the DDF approach used to create the state. DDF is used to construct BRST invariant vertex operators for highly excited string states and their scattering amplitudes. We believe that the origin of the chaotic behavior is to be ascribed to the huge degeneracy of string excitations at a fixed level $N$ and generic spin/helicity $J$ (for $J\ll N$). We suspect -- but are so far unable to prove -- that the relevant mixing matrix, even at lowest non-trivial order in the string coupling $g_s$, will be a random matrix with many separate blocks, one per each spin or representation of the relevant Lorentz group.

The quantum resonant systems analyzed in \cite{Craps:2022ese} involve oscillators that bear striking similarity to the string oscillators that play a crucial role in our analysis as building blocks of highly excited string states at fixed level $N$ but random spin (in principle any $N\ge s\ge J$).

Let us list some of many open questions one can address. (i) What is the detailed mechanism of the onset of chaotic behavior in string theory? (ii) Is there an underlying theory of the observed log-normal distributions or is it simply a convenient model? (iii) In analogy to the level spacings, can one devise an operator whose eigenvalues are the locations of the amplitude's maximum points? (iv) We expect that a similar distribution occur would also for other scattering processes, including classical processes like the pinball problem, QM like the BMN \cite{Berenstein:2002jq} and BFSS models \cite{Fukushima:2022lsd}, QFT like the massive Schwinger model, or scattering in other string theories. This could be explicitly studied. (vi) Performing a similar analysis higher point functions of the DDF string would be an important further study. A first step was taken already in \cite{Hashimoto:2022bll}. (vi) Can one connect the chaotic behavior of string amplitudes with the chaotic behavior in black hole dynamics in view of the string/BH correspondence?

We believe our analysis represents a step forward in the direction of quantifying the chaotic behaviour of string amplitudes -- even simple and calculable ones -- and of identifying its origin in the huge degeneracy of string states obtained combining many spin components by varying slightly the partition of integers at a given large level $N$. This description represents a workable proxy of the superposition of microstates that should capture the quantum dynamics of black holes.


\section*{Acknowledgments}
We would like to thank R.~Benzi, B.~Craps, F.~Popov, D.~Gross, S.~Negro, G.~Parisi, V.~Rosenhaus, and S.~Yankielowicz for useful discussions. The work of M.~B. and M.~F. is partially supported by the MIUR PRIN Grant 2020KR4KN2 ``String Theory as a bridge between Gauge Theories and Quantum Gravity''. The work of J.~S. was supported by a
center of excellence of the Israel Science Foundation (grant number 2289/18). The work of D.~W. was supported by the Quantum Gravity Unit
of the Okinawa Institute of Science and Technology Graduate University (OIST). 

\bibliography{SACS}

\begin{thebibliography}{43}%
\makeatletter
\providecommand \@ifxundefined [1]{%
 \@ifx{#1\undefined}
}%
\providecommand \@ifnum [1]{%
 \ifnum #1\expandafter \@firstoftwo
 \else \expandafter \@secondoftwo
 \fi
}%
\providecommand \@ifx [1]{%
 \ifx #1\expandafter \@firstoftwo
 \else \expandafter \@secondoftwo
 \fi
}%
\providecommand \natexlab [1]{#1}%
\providecommand \enquote  [1]{``#1''}%
\providecommand \bibnamefont  [1]{#1}%
\providecommand \bibfnamefont [1]{#1}%
\providecommand \citenamefont [1]{#1}%
\providecommand \href@noop [0]{\@secondoftwo}%
\providecommand \href [0]{\begingroup \@sanitize@url \@href}%
\providecommand \@href[1]{\@@startlink{#1}\@@href}%
\providecommand \@@href[1]{\endgroup#1\@@endlink}%
\providecommand \@sanitize@url [0]{\catcode `\\12\catcode `\$12\catcode
  `\&12\catcode `\#12\catcode `\^12\catcode `\_12\catcode `\%12\relax}%
\providecommand \@@startlink[1]{}%
\providecommand \@@endlink[0]{}%
\providecommand \url  [0]{\begingroup\@sanitize@url \@url }%
\providecommand \@url [1]{\endgroup\@href {#1}{\urlprefix }}%
\providecommand \urlprefix  [0]{URL }%
\providecommand \Eprint [0]{\href }%
\providecommand \doibase [0]{http://dx.doi.org/}%
\providecommand \selectlanguage [0]{\@gobble}%
\providecommand \bibinfo  [0]{\@secondoftwo}%
\providecommand \bibfield  [0]{\@secondoftwo}%
\providecommand \translation [1]{[#1]}%
\providecommand \BibitemOpen [0]{}%
\providecommand \bibitemStop [0]{}%
\providecommand \bibitemNoStop [0]{.\EOS\space}%
\providecommand \EOS [0]{\spacefactor3000\relax}%
\providecommand \BibitemShut  [1]{\csname bibitem#1\endcsname}%
\let\auto@bib@innerbib\@empty
\bibitem [{\citenamefont {Rosenhaus}(2021)}]{Rosenhaus:2020tmv}%
  \BibitemOpen
  \bibfield  {author} {\bibinfo {author} {\bibfnamefont {V.}~\bibnamefont
  {Rosenhaus}},\ }\href {\doibase 10.1103/PhysRevLett.127.021601} {\bibfield
  {journal} {\bibinfo  {journal} {Phys. Rev. Lett.}\ }\textbf {\bibinfo
  {volume} {127}},\ \bibinfo {pages} {021601} (\bibinfo {year} {2021})},\
  \Eprint {http://arxiv.org/abs/2003.07381} {arXiv:2003.07381 [hep-th]}
  \BibitemShut {NoStop}%
\bibitem [{\citenamefont {Gross}\ and\ \citenamefont
  {Rosenhaus}(2021)}]{Gross:2021gsj}%
  \BibitemOpen
  \bibfield  {author} {\bibinfo {author} {\bibfnamefont {D.~J.}\ \bibnamefont
  {Gross}}\ and\ \bibinfo {author} {\bibfnamefont {V.}~\bibnamefont
  {Rosenhaus}},\ }\href {\doibase 10.1007/JHEP05(2021)048} {\bibfield
  {journal} {\bibinfo  {journal} {JHEP}\ }\textbf {\bibinfo {volume} {05}},\
  \bibinfo {pages} {048} (\bibinfo {year} {2021})},\ \Eprint
  {http://arxiv.org/abs/2103.15301} {arXiv:2103.15301 [hep-th]} \BibitemShut
  {NoStop}%
\bibitem [{\citenamefont {Rosenhaus}(2022)}]{Rosenhaus:2021xhm}%
  \BibitemOpen
  \bibfield  {author} {\bibinfo {author} {\bibfnamefont {V.}~\bibnamefont
  {Rosenhaus}},\ }\href {\doibase 10.1103/PhysRevLett.129.031601} {\bibfield
  {journal} {\bibinfo  {journal} {Phys. Rev. Lett.}\ }\textbf {\bibinfo
  {volume} {129}},\ \bibinfo {pages} {031601} (\bibinfo {year} {2022})},\
  \Eprint {http://arxiv.org/abs/2112.10269} {arXiv:2112.10269 [hep-th]}
  \BibitemShut {NoStop}%
\bibitem [{\citenamefont {Berry}\ and\ \citenamefont
  {Tabor}(1977)}]{Berry:1977wk}%
  \BibitemOpen
  \bibfield  {author} {\bibinfo {author} {\bibfnamefont {M.~V.}\ \bibnamefont
  {Berry}}\ and\ \bibinfo {author} {\bibfnamefont {M.}~\bibnamefont {Tabor}},\
  }\href {\doibase 10.1098/rspa.1977.0140} {\bibfield  {journal} {\bibinfo
  {journal} {Proc. R. Soc. Lond. A}\ }\textbf {\bibinfo {volume} {356}},\
  \bibinfo {pages} {375} (\bibinfo {year} {1977})}\BibitemShut {NoStop}%
\bibitem [{\citenamefont {Oganesyan}\ and\ \citenamefont {Huse}(2007)}]{Huse}%
  \BibitemOpen
  \bibfield  {author} {\bibinfo {author} {\bibfnamefont {V.}~\bibnamefont
  {Oganesyan}}\ and\ \bibinfo {author} {\bibfnamefont {D.~A.}\ \bibnamefont
  {Huse}},\ }\href@noop {} {\bibfield  {journal} {\bibinfo  {journal} {Phys.
  Rev. B}\ }\textbf {\bibinfo {volume} {75}},\ \bibinfo {pages} {155111}
  (\bibinfo {year} {2007})}\BibitemShut {NoStop}%
\bibitem [{\citenamefont {Srdin\v{s}ek}\ \emph {et~al.}(2021)\citenamefont
  {Srdin\v{s}ek}, \citenamefont {Prosen},\ and\ \citenamefont
  {Sotiriadis}}]{Srdinsek:2020bpq}%
  \BibitemOpen
  \bibfield  {author} {\bibinfo {author} {\bibfnamefont {M.}~\bibnamefont
  {Srdin\v{s}ek}}, \bibinfo {author} {\bibfnamefont {T.}~\bibnamefont
  {Prosen}}, \ and\ \bibinfo {author} {\bibfnamefont {S.}~\bibnamefont
  {Sotiriadis}},\ }\href {\doibase 10.1103/PhysRevLett.126.121602} {\bibfield
  {journal} {\bibinfo  {journal} {Phys. Rev. Lett.}\ }\textbf {\bibinfo
  {volume} {126}},\ \bibinfo {pages} {121602} (\bibinfo {year} {2021})},\
  \Eprint {http://arxiv.org/abs/2012.08505} {arXiv:2012.08505
  [cond-mat.stat-mech]} \BibitemShut {NoStop}%
\bibitem [{Note1()}]{Note1}%
  \BibitemOpen
  \bibinfo {note} {In some cases the normalized ratio \(\protect \tilde r_n
  \equiv min\{r_n,\protect \frac {1}{r_n}\}\), defined to be between 0 and 1,
  is used.}\BibitemShut {Stop}%
\bibitem [{\citenamefont {Del~Giudice}\ \emph {et~al.}(1972)\citenamefont
  {Del~Giudice}, \citenamefont {Di~Vecchia},\ and\ \citenamefont
  {Fubini}}]{DelGiudice:1971yjh}%
  \BibitemOpen
  \bibfield  {author} {\bibinfo {author} {\bibfnamefont {E.}~\bibnamefont
  {Del~Giudice}}, \bibinfo {author} {\bibfnamefont {P.}~\bibnamefont
  {Di~Vecchia}}, \ and\ \bibinfo {author} {\bibfnamefont {S.}~\bibnamefont
  {Fubini}},\ }\href {\doibase 10.1016/0003-4916(72)90272-2} {\bibfield
  {journal} {\bibinfo  {journal} {Annals Phys.}\ }\textbf {\bibinfo {volume}
  {70}},\ \bibinfo {pages} {378} (\bibinfo {year} {1972})}\BibitemShut
  {NoStop}%
\bibitem [{\citenamefont {Hindmarsh}\ and\ \citenamefont
  {Skliros}(2011)}]{Hindmarsh:2010if}%
  \BibitemOpen
  \bibfield  {author} {\bibinfo {author} {\bibfnamefont {M.}~\bibnamefont
  {Hindmarsh}}\ and\ \bibinfo {author} {\bibfnamefont {D.}~\bibnamefont
  {Skliros}},\ }\href {\doibase 10.1103/PhysRevLett.106.081602} {\bibfield
  {journal} {\bibinfo  {journal} {Phys. Rev. Lett.}\ }\textbf {\bibinfo
  {volume} {106}},\ \bibinfo {pages} {081602} (\bibinfo {year} {2011})},\
  \Eprint {http://arxiv.org/abs/1006.2559} {arXiv:1006.2559 [hep-th]}
  \BibitemShut {NoStop}%
\bibitem [{\citenamefont {Skliros}\ and\ \citenamefont
  {Hindmarsh}(2011)}]{Skliros:2011si}%
  \BibitemOpen
  \bibfield  {author} {\bibinfo {author} {\bibfnamefont {D.}~\bibnamefont
  {Skliros}}\ and\ \bibinfo {author} {\bibfnamefont {M.}~\bibnamefont
  {Hindmarsh}},\ }\href {\doibase 10.1103/PhysRevD.84.126001} {\bibfield
  {journal} {\bibinfo  {journal} {Phys. Rev. D}\ }\textbf {\bibinfo {volume}
  {84}},\ \bibinfo {pages} {126001} (\bibinfo {year} {2011})},\ \Eprint
  {http://arxiv.org/abs/1107.0730} {arXiv:1107.0730 [hep-th]} \BibitemShut
  {NoStop}%
\bibitem [{\citenamefont {Bianchi}\ and\ \citenamefont
  {Firrotta}(2020)}]{Bianchi:2019ywd}%
  \BibitemOpen
  \bibfield  {author} {\bibinfo {author} {\bibfnamefont {M.}~\bibnamefont
  {Bianchi}}\ and\ \bibinfo {author} {\bibfnamefont {M.}~\bibnamefont
  {Firrotta}},\ }\href {\doibase 10.1016/j.nuclphysb.2020.114943} {\bibfield
  {journal} {\bibinfo  {journal} {Nucl. Phys. B}\ }\textbf {\bibinfo {volume}
  {952}},\ \bibinfo {pages} {114943} (\bibinfo {year} {2020})},\ \Eprint
  {http://arxiv.org/abs/1902.07016} {arXiv:1902.07016 [hep-th]} \BibitemShut
  {NoStop}%
\bibitem [{\citenamefont {Addazi}\ \emph {et~al.}(2021)\citenamefont {Addazi},
  \citenamefont {Bianchi}, \citenamefont {Firrotta},\ and\ \citenamefont
  {Marcian\`o}}]{Addazi:2020obs}%
  \BibitemOpen
  \bibfield  {author} {\bibinfo {author} {\bibfnamefont {A.}~\bibnamefont
  {Addazi}}, \bibinfo {author} {\bibfnamefont {M.}~\bibnamefont {Bianchi}},
  \bibinfo {author} {\bibfnamefont {M.}~\bibnamefont {Firrotta}}, \ and\
  \bibinfo {author} {\bibfnamefont {A.}~\bibnamefont {Marcian\`o}},\ }\href
  {\doibase 10.1016/j.nuclphysb.2021.115356} {\bibfield  {journal} {\bibinfo
  {journal} {Nucl. Phys. B}\ }\textbf {\bibinfo {volume} {965}},\ \bibinfo
  {pages} {115356} (\bibinfo {year} {2021})},\ \Eprint
  {http://arxiv.org/abs/2008.02206} {arXiv:2008.02206 [hep-th]} \BibitemShut
  {NoStop}%
\bibitem [{\citenamefont {Aldi}\ and\ \citenamefont
  {Firrotta}(2020)}]{Aldi:2019osr}%
  \BibitemOpen
  \bibfield  {author} {\bibinfo {author} {\bibfnamefont {A.}~\bibnamefont
  {Aldi}}\ and\ \bibinfo {author} {\bibfnamefont {M.}~\bibnamefont
  {Firrotta}},\ }\href {\doibase 10.1016/j.nuclphysb.2020.115050} {\bibfield
  {journal} {\bibinfo  {journal} {Nucl. Phys. B}\ }\textbf {\bibinfo {volume}
  {955}},\ \bibinfo {pages} {115050} (\bibinfo {year} {2020})},\ \Eprint
  {http://arxiv.org/abs/1912.06177} {arXiv:1912.06177 [hep-th]} \BibitemShut
  {NoStop}%
\bibitem [{\citenamefont {Berry}(1986)}]{Berry:1986}%
  \BibitemOpen
  \bibfield  {author} {\bibinfo {author} {\bibfnamefont {M.~V.}\ \bibnamefont
  {Berry}},\ }in\ \href@noop {} {\emph {\bibinfo {booktitle} {Quantum Chaos and
  Statistical Nuclear Physics}}},\ \bibinfo {editor} {edited by\ \bibinfo
  {editor} {\bibfnamefont {T.~H.}\ \bibnamefont {Seligman}}\ and\ \bibinfo
  {editor} {\bibfnamefont {H.}~\bibnamefont {Nishioka}}}\ (\bibinfo
  {publisher} {Springer Berlin Heidelberg},\ \bibinfo {address} {Berlin,
  Heidelberg},\ \bibinfo {year} {1986})\ pp.\ \bibinfo {pages}
  {1--17}\BibitemShut {NoStop}%
\bibitem [{\citenamefont {Odlyzko}(1987)}]{Odlyzko:1987}%
  \BibitemOpen
  \bibfield  {author} {\bibinfo {author} {\bibfnamefont {A.~M.}\ \bibnamefont
  {Odlyzko}},\ }\href {\doibase 10.1090/s0025-5718-1987-0866115-0} {\bibfield
  {journal} {\bibinfo  {journal} {Mathematics of Computation}\ }\textbf
  {\bibinfo {volume} {48}},\ \bibinfo {pages} {273} (\bibinfo {year}
  {1987})}\BibitemShut {NoStop}%
\bibitem [{\citenamefont {Wolf}(2020)}]{Wolf:2014ulr}%
  \BibitemOpen
  \bibfield  {author} {\bibinfo {author} {\bibfnamefont {M.}~\bibnamefont
  {Wolf}},\ }\href {\doibase 10.1088/1361-6633/ab3de7} {\bibfield  {journal}
  {\bibinfo  {journal} {Rept. Prog. Phys.}\ }\textbf {\bibinfo {volume} {83}},\
  \bibinfo {pages} {036001} (\bibinfo {year} {2020})},\ \Eprint
  {http://arxiv.org/abs/1410.1214} {arXiv:1410.1214 [math-ph]} \BibitemShut
  {NoStop}%
\bibitem [{\citenamefont {Gutzwiller}(1983)}]{Gutzwiller:1983}%
  \BibitemOpen
  \bibfield  {author} {\bibinfo {author} {\bibfnamefont {M.~C.}\ \bibnamefont
  {Gutzwiller}},\ }\href {\doibase
  https://doi.org/10.1016/0167-2789(83)90138-0} {\bibfield  {journal} {\bibinfo
   {journal} {Physica D: Nonlinear Phenomena}\ }\textbf {\bibinfo {volume}
  {7}},\ \bibinfo {pages} {341} (\bibinfo {year} {1983})}\BibitemShut {NoStop}%
\bibitem [{Note2()}]{Note2}%
  \BibitemOpen
  \bibinfo {note} {A relation to string amplitudes is also found in the
  representation of the Veneziano amplitude in terms of zeta functions \cite
  {He:2015jla,CastroPerelman:2022ypp}. However, the Veneziano amplitude is
  certainly not chaotic, so further steps are necessary to link chaos in the
  zeta function to that of string amplitudes.}\BibitemShut {Stop}%
\bibitem [{Note3()}]{Note3}%
  \BibitemOpen
  \bibinfo {note} {As opposed to the trivial zeros located at \(s = -2n\) for
  integer \(n\).}\BibitemShut {Stop}%
\bibitem [{Note4()}]{Note4}%
  \BibitemOpen
  \bibinfo {note} {Since \(\zeta (s^*) = \zeta ^*(s)\) we can discuss zeros
  with positive imaginary part only.}\BibitemShut {Stop}%
\bibitem [{Note5()}]{Note5}%
  \BibitemOpen
  \bibinfo {note} {The logarithmic normalization is not essential in this case.
  One can normalize also by dividing by the mean, \(\protect \bar \delta _n
  \equiv \delta _n/\langle \delta \rangle \). In both cases \(\protect \bar
  \delta =1\).}\BibitemShut {Stop}%
\bibitem [{Note6()}]{Note6}%
  \BibitemOpen
  \bibinfo {note} {More precisely, we write the PDF for $2\times 2$ random
  matrices, which is generally considered an excellent approximation of the PDF
  of spacings for large matrices as well.}\BibitemShut {Stop}%
\bibitem [{Note7()}]{Note7}%
  \BibitemOpen
  \bibinfo {note} {This specific value of \(\sigma \) is the one for which the
  relative entropy, \(\DOTSI \intop \ilimits@ f_{\protect \text {GUE}}\protect
  \qopname \relax o{log}(f_{\protect \text {GUE}}/f_{\protect \text {LN}})\),
  is minimized, with the minimum being \(\approx 0.008\).}\BibitemShut {Stop}%
\bibitem [{\citenamefont {Odlyzko}()}]{Odlyzko:Zeta}%
  \BibitemOpen
  \bibfield  {author} {\bibinfo {author} {\bibfnamefont {A.}~\bibnamefont
  {Odlyzko}},\ }\href@noop {} {\enquote {\bibinfo {title} {Tables of zeros of
  the riemann zeta function},}\ }\bibinfo {howpublished}
  {\url{http://www.dtc.umn.edu/~odlyzko/zeta_tables/}},\ \bibinfo {note}
  {accessed: June 2022}\BibitemShut {NoStop}%
\bibitem [{Note8()}]{Note8}%
  \BibitemOpen
  \bibinfo {note} {We implicitly set the intrinsic length scale in the problem
  to one when writing the metric.}\BibitemShut {Stop}%
\bibitem [{\citenamefont {Bianchi}\ and\ \citenamefont
  {Lopez}(2010)}]{Bianchi:2010dy}%
  \BibitemOpen
  \bibfield  {author} {\bibinfo {author} {\bibfnamefont {M.}~\bibnamefont
  {Bianchi}}\ and\ \bibinfo {author} {\bibfnamefont {L.}~\bibnamefont
  {Lopez}},\ }\href {\doibase 10.1007/JHEP07(2010)065} {\bibfield  {journal}
  {\bibinfo  {journal} {JHEP}\ }\textbf {\bibinfo {volume} {07}},\ \bibinfo
  {pages} {065} (\bibinfo {year} {2010})},\ \Eprint
  {http://arxiv.org/abs/1002.3058} {arXiv:1002.3058 [hep-th]} \BibitemShut
  {NoStop}%
\bibitem [{\citenamefont {Bianchi}\ \emph {et~al.}(2011)\citenamefont
  {Bianchi}, \citenamefont {Lopez},\ and\ \citenamefont
  {Richter}}]{Bianchi:2010es}%
  \BibitemOpen
  \bibfield  {author} {\bibinfo {author} {\bibfnamefont {M.}~\bibnamefont
  {Bianchi}}, \bibinfo {author} {\bibfnamefont {L.}~\bibnamefont {Lopez}}, \
  and\ \bibinfo {author} {\bibfnamefont {R.}~\bibnamefont {Richter}},\ }\href
  {\doibase 10.1007/JHEP03(2011)051} {\bibfield  {journal} {\bibinfo  {journal}
  {JHEP}\ }\textbf {\bibinfo {volume} {03}},\ \bibinfo {pages} {051} (\bibinfo
  {year} {2011})},\ \Eprint {http://arxiv.org/abs/1010.1177} {arXiv:1010.1177
  [hep-th]} \BibitemShut {NoStop}%
\bibitem [{\citenamefont {Bianchi}\ and\ \citenamefont
  {Teresi}(2012)}]{Bianchi:2011se}%
  \BibitemOpen
  \bibfield  {author} {\bibinfo {author} {\bibfnamefont {M.}~\bibnamefont
  {Bianchi}}\ and\ \bibinfo {author} {\bibfnamefont {P.}~\bibnamefont
  {Teresi}},\ }\href {\doibase 10.1007/JHEP01(2012)161} {\bibfield  {journal}
  {\bibinfo  {journal} {JHEP}\ }\textbf {\bibinfo {volume} {01}},\ \bibinfo
  {pages} {161} (\bibinfo {year} {2012})},\ \Eprint
  {http://arxiv.org/abs/1108.1071} {arXiv:1108.1071 [hep-th]} \BibitemShut
  {NoStop}%
\bibitem [{\citenamefont {Black}\ and\ \citenamefont
  {Monni}(2012)}]{Black:2011ep}%
  \BibitemOpen
  \bibfield  {author} {\bibinfo {author} {\bibfnamefont {W.}~\bibnamefont
  {Black}}\ and\ \bibinfo {author} {\bibfnamefont {C.}~\bibnamefont {Monni}},\
  }\href {\doibase 10.1016/j.nuclphysb.2012.02.009} {\bibfield  {journal}
  {\bibinfo  {journal} {Nucl. Phys. B}\ }\textbf {\bibinfo {volume} {859}},\
  \bibinfo {pages} {299} (\bibinfo {year} {2012})},\ \Eprint
  {http://arxiv.org/abs/1107.4321} {arXiv:1107.4321 [hep-th]} \BibitemShut
  {NoStop}%
\bibitem [{\citenamefont {Schlotterer}(2011)}]{Schlotterer:2010kk}%
  \BibitemOpen
  \bibfield  {author} {\bibinfo {author} {\bibfnamefont {O.}~\bibnamefont
  {Schlotterer}},\ }\href {\doibase 10.1016/j.nuclphysb.2011.03.026} {\bibfield
   {journal} {\bibinfo  {journal} {Nucl. Phys. B}\ }\textbf {\bibinfo {volume}
  {849}},\ \bibinfo {pages} {433} (\bibinfo {year} {2011})},\ \Eprint
  {http://arxiv.org/abs/1011.1235} {arXiv:1011.1235 [hep-th]} \BibitemShut
  {NoStop}%
\bibitem [{\citenamefont {Aldi}\ \emph {et~al.}(2021)\citenamefont {Aldi},
  \citenamefont {Bianchi},\ and\ \citenamefont {Firrotta}}]{Aldi:2020qfu}%
  \BibitemOpen
  \bibfield  {author} {\bibinfo {author} {\bibfnamefont {A.}~\bibnamefont
  {Aldi}}, \bibinfo {author} {\bibfnamefont {M.}~\bibnamefont {Bianchi}}, \
  and\ \bibinfo {author} {\bibfnamefont {M.}~\bibnamefont {Firrotta}},\ }\href
  {\doibase 10.1016/j.physletb.2020.136037} {\bibfield  {journal} {\bibinfo
  {journal} {Phys. Lett. B}\ }\textbf {\bibinfo {volume} {813}},\ \bibinfo
  {pages} {136037} (\bibinfo {year} {2021})},\ \Eprint
  {http://arxiv.org/abs/2010.04082} {arXiv:2010.04082 [hep-th]} \BibitemShut
  {NoStop}%
\bibitem [{\citenamefont {Aldi}\ \emph {et~al.}(2022)\citenamefont {Aldi},
  \citenamefont {Bianchi},\ and\ \citenamefont {Firrotta}}]{Aldi:2021zhh}%
  \BibitemOpen
  \bibfield  {author} {\bibinfo {author} {\bibfnamefont {A.}~\bibnamefont
  {Aldi}}, \bibinfo {author} {\bibfnamefont {M.}~\bibnamefont {Bianchi}}, \
  and\ \bibinfo {author} {\bibfnamefont {M.}~\bibnamefont {Firrotta}},\ }\href
  {\doibase 10.1016/j.nuclphysb.2021.115625} {\bibfield  {journal} {\bibinfo
  {journal} {Nucl. Phys. B}\ }\textbf {\bibinfo {volume} {974}},\ \bibinfo
  {pages} {115625} (\bibinfo {year} {2022})},\ \Eprint
  {http://arxiv.org/abs/2101.07054} {arXiv:2101.07054 [hep-th]} \BibitemShut
  {NoStop}%
\bibitem [{\citenamefont {Firrotta}\ and\ \citenamefont
  {Rosenhaus}(2022)}]{Firrotta:2022cku}%
  \BibitemOpen
  \bibfield  {author} {\bibinfo {author} {\bibfnamefont {M.}~\bibnamefont
  {Firrotta}}\ and\ \bibinfo {author} {\bibfnamefont {V.}~\bibnamefont
  {Rosenhaus}},\ }\href {\doibase 10.1007/JHEP09(2022)211} {\bibfield
  {journal} {\bibinfo  {journal} {JHEP}\ }\textbf {\bibinfo {volume} {09}},\
  \bibinfo {pages} {211} (\bibinfo {year} {2022})},\ \Eprint
  {http://arxiv.org/abs/2207.01641} {arXiv:2207.01641 [hep-th]} \BibitemShut
  {NoStop}%
\bibitem [{Note9()}]{Note9}%
  \BibitemOpen
  \bibinfo {note} {More precisely, in terms of the $|{\protect \cal A}|$ there
  are only maxima. In this sense, unlike the case of the leaky torus, all the
  zeros of the derivative represent peaks.}\BibitemShut {Stop}%
\bibitem [{\citenamefont {Erd\"{o}s}\ and\ \citenamefont
  {Lehner}(1941)}]{Erdos:1941}%
  \BibitemOpen
  \bibfield  {author} {\bibinfo {author} {\bibfnamefont {P.}~\bibnamefont
  {Erd\"{o}s}}\ and\ \bibinfo {author} {\bibfnamefont {J.}~\bibnamefont
  {Lehner}},\ }\href {http://projecteuclid.org/euclid.dmj/1077492649}
  {\bibfield  {journal} {\bibinfo  {journal} {Duke Math. J.}\ }\textbf
  {\bibinfo {volume} {8}},\ \bibinfo {pages} {335} (\bibinfo {year}
  {1941})}\BibitemShut {NoStop}%
\bibitem [{Note10()}]{Note10}%
  \BibitemOpen
  \bibinfo {note} {More precisely the distribution is peaked around \( J
  \approx \protect \frac {1}{C} \protect \sqrt {N}\protect \qopname \relax
  o{log}N\) with the same constant \(C\) as in \protect \textup {\hbox
  {\mathsurround \z@ \protect \normalfont (\ignorespaces \ref
  {eq:partitions}\unskip \@@italiccorr )}}.}\BibitemShut {Stop}%
\bibitem [{Note11()}]{Note11}%
  \BibitemOpen
  \bibinfo {note} {The form of each term \(\protect \qopname \relax o{sin}(\pi
  m \protect \qopname \relax o{cos}^2\protect \frac \alpha 2)\) is obtained
  from a ratio of \(\Gamma \) functions, after a Stirling approximation
  assuming \(m\) is large, which is not always the case.}\BibitemShut {Stop}%
\bibitem [{\citenamefont {Craps}\ \emph {et~al.}(2022)\citenamefont {Craps},
  \citenamefont {De~Clerck}, \citenamefont {Evnin}, \citenamefont {Hacker},\
  and\ \citenamefont {Pavlov}}]{Craps:2022ese}%
  \BibitemOpen
  \bibfield  {author} {\bibinfo {author} {\bibfnamefont {B.}~\bibnamefont
  {Craps}}, \bibinfo {author} {\bibfnamefont {M.}~\bibnamefont {De~Clerck}},
  \bibinfo {author} {\bibfnamefont {O.}~\bibnamefont {Evnin}}, \bibinfo
  {author} {\bibfnamefont {P.}~\bibnamefont {Hacker}}, \ and\ \bibinfo {author}
  {\bibfnamefont {M.}~\bibnamefont {Pavlov}},\ }\href {\doibase
  10.21468/SciPostPhys.13.4.090} {\bibfield  {journal} {\bibinfo  {journal}
  {SciPost Phys.}\ }\textbf {\bibinfo {volume} {13}},\ \bibinfo {pages} {090}
  (\bibinfo {year} {2022})},\ \Eprint {http://arxiv.org/abs/2202.13924}
  {arXiv:2202.13924 [quant-ph]} \BibitemShut {NoStop}%
\bibitem [{\citenamefont {Berenstein}\ \emph {et~al.}(2002)\citenamefont
  {Berenstein}, \citenamefont {Maldacena},\ and\ \citenamefont
  {Nastase}}]{Berenstein:2002jq}%
  \BibitemOpen
  \bibfield  {author} {\bibinfo {author} {\bibfnamefont {D.~E.}\ \bibnamefont
  {Berenstein}}, \bibinfo {author} {\bibfnamefont {J.~M.}\ \bibnamefont
  {Maldacena}}, \ and\ \bibinfo {author} {\bibfnamefont {H.~S.}\ \bibnamefont
  {Nastase}},\ }\href {\doibase 10.1088/1126-6708/2002/04/013} {\bibfield
  {journal} {\bibinfo  {journal} {JHEP}\ }\textbf {\bibinfo {volume} {04}},\
  \bibinfo {pages} {013} (\bibinfo {year} {2002})},\ \Eprint
  {http://arxiv.org/abs/hep-th/0202021} {arXiv:hep-th/0202021} \BibitemShut
  {NoStop}%
\bibitem [{\citenamefont {Fukushima}\ and\ \citenamefont
  {Yoshida}(2022)}]{Fukushima:2022lsd}%
  \BibitemOpen
  \bibfield  {author} {\bibinfo {author} {\bibfnamefont {O.}~\bibnamefont
  {Fukushima}}\ and\ \bibinfo {author} {\bibfnamefont {K.}~\bibnamefont
  {Yoshida}},\ }\href {\doibase 10.1007/JHEP09(2022)039} {\bibfield  {journal}
  {\bibinfo  {journal} {JHEP}\ }\textbf {\bibinfo {volume} {09}},\ \bibinfo
  {pages} {039} (\bibinfo {year} {2022})},\ \Eprint
  {http://arxiv.org/abs/2204.06391} {arXiv:2204.06391 [hep-th]} \BibitemShut
  {NoStop}%
\bibitem [{\citenamefont {Hashimoto}\ \emph {et~al.}(2022)\citenamefont
  {Hashimoto}, \citenamefont {Matsuo},\ and\ \citenamefont
  {Yoda}}]{Hashimoto:2022bll}%
  \BibitemOpen
  \bibfield  {author} {\bibinfo {author} {\bibfnamefont {K.}~\bibnamefont
  {Hashimoto}}, \bibinfo {author} {\bibfnamefont {Y.}~\bibnamefont {Matsuo}}, \
  and\ \bibinfo {author} {\bibfnamefont {T.}~\bibnamefont {Yoda}},\ }\href
  {\doibase 10.1007/JHEP11(2022)147} {\bibfield  {journal} {\bibinfo  {journal}
  {JHEP}\ }\textbf {\bibinfo {volume} {11}},\ \bibinfo {pages} {147} (\bibinfo
  {year} {2022})},\ \Eprint {http://arxiv.org/abs/2208.08380} {arXiv:2208.08380
  [hep-th]} \BibitemShut {NoStop}%
\bibitem [{\citenamefont {He}\ \emph {et~al.}(2016)\citenamefont {He},
  \citenamefont {Jejjala},\ and\ \citenamefont {Minic}}]{He:2015jla}%
  \BibitemOpen
  \bibfield  {author} {\bibinfo {author} {\bibfnamefont {Y.-H.}\ \bibnamefont
  {He}}, \bibinfo {author} {\bibfnamefont {V.}~\bibnamefont {Jejjala}}, \ and\
  \bibinfo {author} {\bibfnamefont {D.}~\bibnamefont {Minic}},\ }\href
  {\doibase 10.1142/S0217751X16502018} {\bibfield  {journal} {\bibinfo
  {journal} {Int. J. Mod. Phys. A}\ }\textbf {\bibinfo {volume} {31}},\
  \bibinfo {pages} {1650201} (\bibinfo {year} {2016})},\ \Eprint
  {http://arxiv.org/abs/1501.01975} {arXiv:1501.01975 [hep-th]} \BibitemShut
  {NoStop}%
\bibitem [{\citenamefont {Castro~Perelman}(2022)}]{CastroPerelman:2022ypp}%
  \BibitemOpen
  \bibfield  {author} {\bibinfo {author} {\bibfnamefont {C.}~\bibnamefont
  {Castro~Perelman}},\ }\href {\doibase 10.1140/epjc/s10052-022-10429-3}
  {\bibfield  {journal} {\bibinfo  {journal} {Eur. Phys. J. C}\ }\textbf
  {\bibinfo {volume} {82}},\ \bibinfo {pages} {469} (\bibinfo {year}
  {2022})}\BibitemShut {NoStop}%
\end{thebibliography}%

\end{document}